\documentclass[]{aastex631}
\usepackage{natbib}
\usepackage{amsmath}
\usepackage{comment}
\usepackage{color}

\shorttitle{CMEs and Dimmings}
\shortauthors{Jin et al.}

\begin{document}
\title{Coronal Mass Ejections and Dimmings: A Comparative Study using MHD Simulations and SDO Observations}

\correspondingauthor{Meng Jin}
\email{jinmeng@lmsal.com}

\author[0000-0002-9672-3873]{Meng Jin}
\affiliation{Lockheed Martin Solar and Astrophysics Lab (LMSAL), Palo Alto, CA 94304, USA}
\affiliation{SETI Institute, Mountain View, CA 94043, USA}

\author[0000-0003-2110-9753]{Mark C. M. Cheung}
\affiliation{Lockheed Martin Solar and Astrophysics Lab (LMSAL), Palo Alto, CA 94304, USA}

\author[0000-0002-6338-0691]{Marc L. DeRosa}
\affiliation{Lockheed Martin Solar and Astrophysics Lab (LMSAL), Palo Alto, CA 94304, USA}

\author[0000-0001-6119-0221]{Nariaki V. Nitta}
\affiliation{Lockheed Martin Solar and Astrophysics Lab (LMSAL), Palo Alto, CA 94304, USA}

\author[0000-0002-6010-8182]{Carolus J. Schrijver}
\affiliation{Lockheed Martin Solar and Astrophysics Lab (LMSAL), Palo Alto, CA 94304, USA}

\begin{abstract}
Solar coronal dimmings have been observed extensively in the past two decades. Due to their close association with coronal mass ejections (CMEs), there is a critical need to improve our understanding of the physical processes that cause dimmings as well as their relationship with CMEs. In this study, we investigate coronal dimmings by combining simulation and observational efforts. By utilizing a data-constrained global magnetohydrodynamics model (AWSoM: Alfv\'{e}n-wave Solar Model), we simulate coronal dimmings resulting from different CME energetics and flux rope configurations. We synthesize the emissions of different EUV spectral bands/lines and compare with SDO/AIA and EVE observations. A detailed analysis of the simulation and observation data suggests that the transient dimming / brightening are related to plasma heating processes, while the long-lasting core and remote dimmings are caused by mass loss process induced by the CME. Moreover, the interaction between the erupting flux rope with different orientations and the global solar corona could significantly influence the coronal dimming patterns. Using metrics such as dimming depth and dimming slope, we investigate the relationship between dimmings and CME properties (e.g., CME mass, CME speed) in the simulation. Our result suggests that coronal dimmings encode important information about the associated CMEs, which provides a physical basis for detecting stellar CMEs from distant solar-like stars. 
\end{abstract}


\section{Introduction}
``Coronal dimming" refers to the reduction in intensity on or near the solar disk across a large area during solar eruptive events. It was first observed in white light corona and described as a ``depletion" \citep{hansen74} and later was found in solar X-ray observations as ``transient coronal holes" \citep{rust76}. Interest in coronal dimmings increased after they were identified in images from multiple EUV channels with different emission temperatures \citep{thompson98} by Solar and Heliospheric Observatory (SOHO)/Extreme-ultraviolet Imaging (EIT) observations \citep{dela95}. Coronal dimmings were often found to be associated with coronal EUV waves (also called ``EIT waves", \citealt{thompson99}). EUV observations at higher temporal ($\sim$12 s) and spatial resolution ($\sim$1.2 arcsec) in seven channels from the Solar Dynamics Observatory (SDO; \citealt{pesnell12})/Atmospheric Imaging Assembly (AIA; \citealt{lemen12}) as well as the high spectral resolution data from SDO/Extreme Ultraviolet Variability Experiment (EVE; \citealt{woods12}) provided unprecedented opportunities to study coronal dimmings \citep{mason14, mason16}.

Two decades of solar observations suggest that the majority of coronal dimmings are associated with coronal mass ejections (CMEs; e.g., \citealt{sterling97, reinard08}). Furthermore, since observations show simultaneous and co-spatial dimming in multiple coronal lines (e.g., \citealt{zarro99, sterling00}) and the spectroscopic observations show that the dimming region has up-flowing expanding plasma (e.g., \citealt{harra01, harra07, imada07, jin09, attrill10, tian12}), it is widely accepted that coronal dimmings are due to plasma evacuation during the launch of CMEs. Specifically, dimming areas are believed to correspond to the footprint of the erupting flux rope. This physical picture is also supported by  magnetohydrodynamics (MHD) modeling results (e.g., \citealt{cohen09,downs12}).

In addition, coronal dimmings are closely correlated with the observed physical properties of the associated CMEs, including mass, speed, and energy (e.g., \citealt{hudson96, sterling97, harrison03, zhukov04, asch09, cheng16, krista17}). In this regard, coronal dimmings provide useful information for space weather forecasts. For example, \citet{krista13} found correlations between the magnitudes of dimmings/flares and CME masses by studying the variations between the recurring eruptions and dimmings. Recently, by using SDO/EVE observations, \citet{mason16} found linear relationships between the speeds and masses of CMEs and coronal dimming properties (e.g., dimming depth and dimming slope), and suggested their relationships could be used for space weather operations of estimating CME mass and speed. Using logarithmic base-ratio images of SDO/AIA to measure the time-integrated coronal dimming parameters (e.g., the size of dimming region, total unsigned magnetic flux, total brightness) in 62 events during 2010-2012, \citet{dissauer18, dissauer19} found that CME mass is strongly correlated with coronal dimming parameters. In addition, as the core dimming is believed to be the footpoints of the erupting flux rope, the enclosed magnetic flux in the dimming region provides an estimate of the total magnetic flux in the erupting flux rope, which has been shown to be consistent with the in-situ flux rope measurements at 1 AU in multiple Earth-directing events \citep{webb00, qiu07}. Since routine remote sensing measurements of the magnetic field strength and geometry of coronal flux ropes remain unavailable, coronal dimmings provide much needed diagnostics that are critical for space weather prediction. Furthermore, by exploring the characteristics of 42 X-class solar flares, \citet{harra16} found that coronal dimmings are the only  signature that could differentiate powerful flares that have CMEs from those that do not. Therefore, dimmings might be one of the best candidate proxies for detection of stellar CMEs from distant Sun-like stars. To gain a better understanding about solar coronal dimmings would provide important reference for solar-stellar connection studies as well as for planning future missions to detect stellar CMEs. 

By modeling a realistic CME event on 2011 February 15 \citep{schrijver11, jin16}, we investigate the coronal dimming phenomenon in a global perspective by combining observational and simulation efforts, in which both the core dimmings and secondary/remote dimmings (e.g., \citealt{thompson00}) are studied. Also, the relationship between coronal dimmings and CMEs in the simulations is investigated and compared with observational results. In \S 2 we briefly introduce the global MHD model utilized in this study. In \S 3 we present the dimming observations for the 2011 February 15 event. The main results are presented in \S 4, followed by discussion and conclusions in \S 5. 

\section{Global Corona \& CME Models}
In this study, Alfv\'{e}n Wave Solar Model \citep{bart14} developed within the Space Weather Modeling Framework (SWMF; \citealt{toth12}) is used to reconstruct the global corona and solar wind environment. With inner boundary condition specified by observed magnetic maps, AWSoM's simulation domain starts from the upper chromosphere, and extends to the corona and heliosphere. The global magnetic map at the inner boundary is taken from the surface-flux transport model of \cite{schrijver03}, into which SDO/HMI magnetogram data within 60$^{\circ}$ from disk center are assimilated. The global magentic map is used as input without any scaling factor applied. Physical processes included in the AWSoM model are multi-species thermodynamics, electron heat conduction (both collisional and collisionless formulations), optically thin radiative cooling, and Alfv\'{e}n-wave turbulence that accelerates and heats the solar wind. The Alfv\'{e}n-wave description is physically self-consistent, including non-Wentzel-Kramers-Brillouin (non-WKB) reflection and physics-based apportioning of turbulent dissipative heating to both electrons and protons. AWSoM has demonstrated the capability to reproduce high-fidelity solar corona and wind conditions \citep{sokolov13, bart14, oran13, oran15, jin16, jin17a}. We also acknowledge one limitation of the model, specifically that, due to the fixed density and temperature at the inner boundary, the chromospheric evaporation process could not be correctly captured that impacts the post-flare emission in the model, which we discuss in more detail in \S 4.2.

The CME is initiated by inserting an analytical Gibson-Low (GL) flux rope \citep{gibson98} into the steady-state global corona and solar wind solution. The GL flux rope has been successfully used before to model CMEs (e.g., \citealt{chip04a, chip04b, lugaz05, chip14, jin16, jin17a}). \citet{jin17b} developed a module (EEGGL) to calculate the GL flux rope parameters based on near-Sun observations to automate the flux rope initiation process in the model. The mathematical form of GL flux rope is derived by solving the magnetohydrostatic equation $(\nabla\times{\bf B})\times{\bf B}-\nabla p-\rho {\bf g}=0$ under solenoidal condition $\nabla\cdot{\bf B}=0$. After inserting the flux rope into the steady-state solar corona solution: i.e. $\rho=\rho_{0}+\rho_{\rm GL}$, ${\bf B = B_{0}+B_{\rm GL}}$, $p=p_{0}+p_{\rm GL}$, and starting the simulation forward in time, the flux rope erupts immediately due to the force imbalance of the combined background--flux-rope system. The simulation setup in this study is similar to that in our previous work \citep{jin16}, to which the reader may refer for further details about the model setup.

\section{Coronal Dimming Observation on 2011 February 15}

\begin{figure}[ht]
\begin{center}$
\begin{array}{c}
\includegraphics[width=0.9\textwidth]{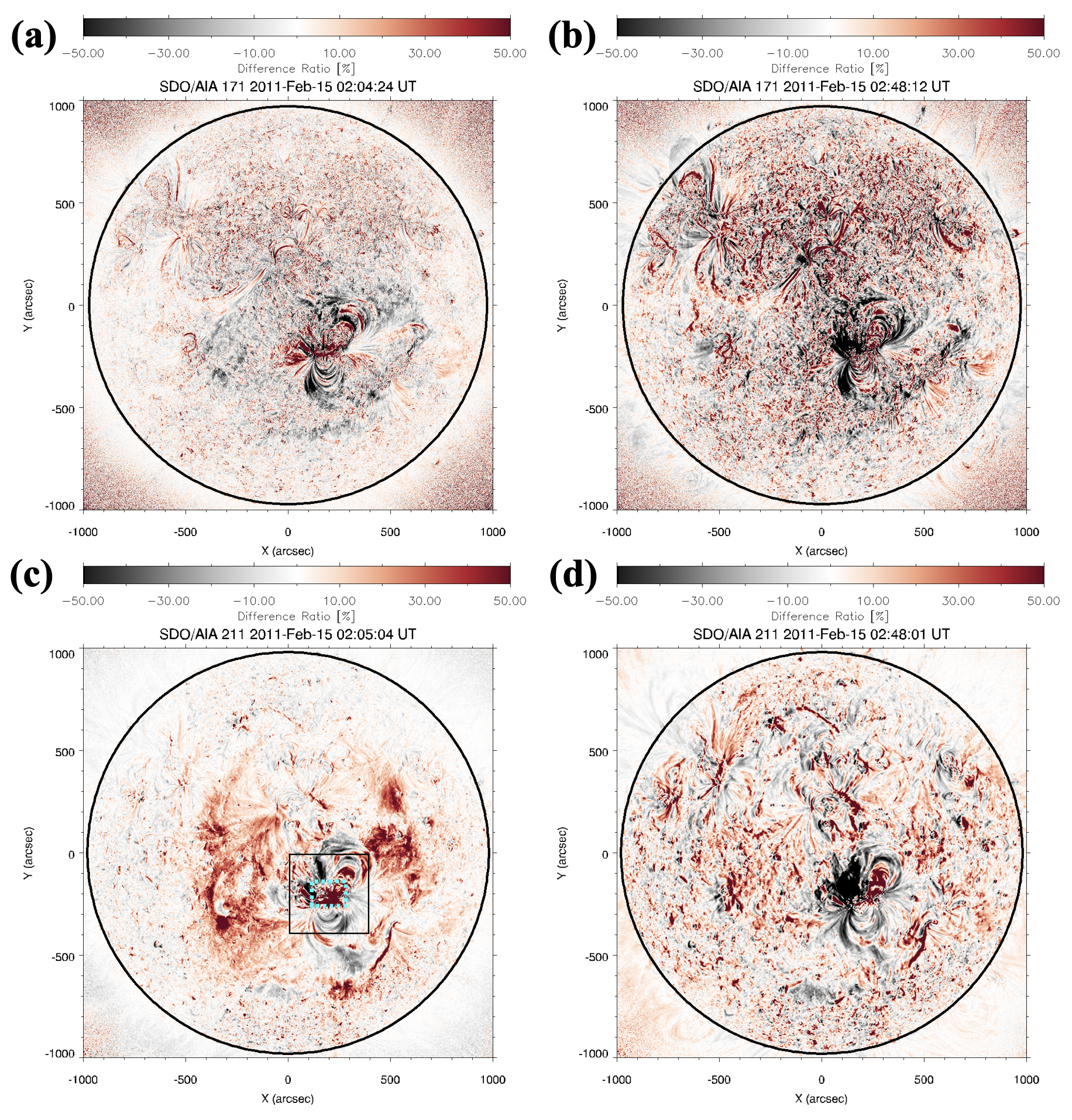}
\end{array}$
\end{center}
\caption{The coronal dimming/brightening observed by SDO/AIA during the 2011 February 15 flare/CME event. Panel (a) and (b) show AIA 171~\AA~percentage base difference image at 02:04:24 UT and 02:48:12 UT (849 and 3477 seconds from the base image at 01:50:15 UT). Panel (c) and (d) show AIA 211~\AA~percentage base difference image at 02:05:04 UT and 02:48:01 UT (902 and 3479 seconds from the base image at 01:50:02 UT). The solid and dashed boxes in panel (c) show the regions used to remove the post-flare loop emissions from synthetic EVE \ion{Fe}{14} 211~\AA~line intensity (see \S 4.2 for details). An animation of this figure is available in the online version of this article, which shows the AIA percentage base difference movie of 171~\AA~and 211~\AA~ from 01:50 UT to 02:48 UT with temporal resolution of $\sim$60 seconds.}
\label{aia_obs}
\end{figure}

The coronal dimming we will focus on is associated with an X2.2-class flare that occurred on 2011 February 15 01:46:50 UT from AR 11158 (see \citealt{schrijver11} for a detailed study about the event). In Figure~\ref{aia_obs}, we show the coronal dimming observation in 2011 February 15 event at two different times and in two of the AIA channels (171~\AA\ and 211~\AA). The two channels are selected to represent the coronal plasma both at transition region ($\log_{10} T/K \sim 5.8$) and coronal ($\log_{10} T/K \sim 6.3$) temperatures. To better identify the dimming signature across the disk, we process the images by the percentage base-difference method, in which base images at 2011 February~15 01:50:15~UT (171 \AA) and 01:50:02~UT (211 \AA) are subtracted. We also correct the solar rotational effect before subtracting. Figures~\ref{aia_obs}(a) and \ref{aia_obs}(c) show the early-stage evolution ($\sim$20 minutes after flare onset) of the waves and dimmings associated with the eruption. The EUV waves in AIA 171~\AA\ show a ``darkening" feature while in AIA 211~\AA\ they are ``brightenings". This suggests a typical plasma warming process \citep{schrijver11, nitta13, liu18}, in which the lower temperature plasmas are heated up to higher temperature by processes like adiabatic compression. For the source region, both channels show evident dimmings that we ascribe to plasma depletion due to the CME. Figures~\ref{aia_obs}(b) and \ref{aia_obs}(d) show the evolution at a later phase ($\sim$60 minutes after flare onset), at which time the waves have traversed the visible disk and more dimming features are evident. Compared with the early-stage phase, the core-dimming is more prominent in both channels and the dimmings at more distant locations are stating to occur. 

\begin{figure}[ht]
\begin{center}$
\begin{array}{c}
\includegraphics[width=1.0\textwidth]{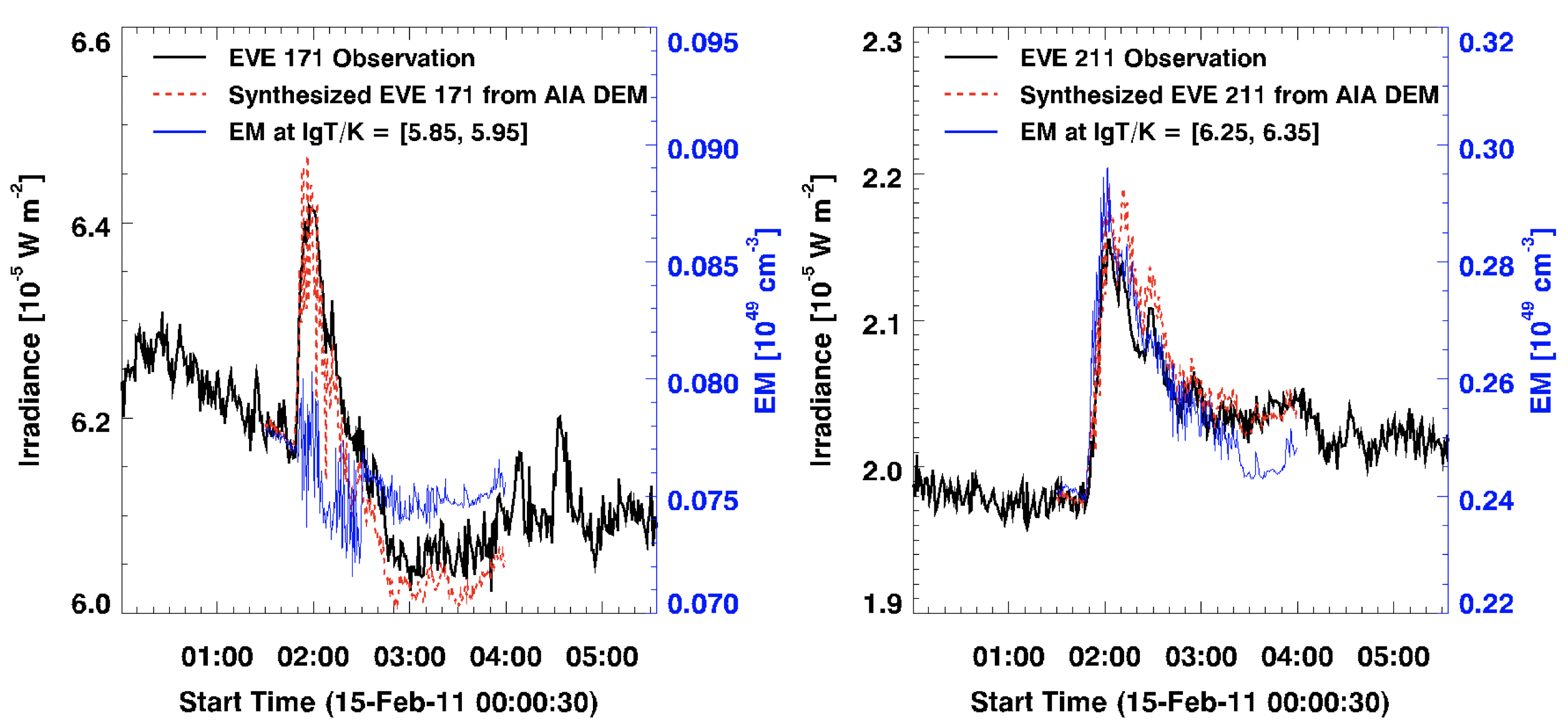}
\end{array}$
\end{center}
\caption{Comparison between the EVE observation (black) and synthesized EVE line intensity from AIA DEM inversion (red). The blue lines show the integrated EM around the peak emission temperatures of the two EVE lines (\ion{Fe}{9} 171 \AA~and \ion{Fe}{14} 211 \AA).}
\label{eve_obs}
\end{figure}

In addition to AIA observations, we plot the EVE light curves of \ion{Fe}{9} 171~\AA\ and \ion{Fe}{14} 211~\AA\ for the event in Figure~\ref{eve_obs}. The EVE \ion{Fe}{9} 171~\AA\ intensity profile contains a significant dimming feature starting at $\sim$02:30:00~UT and has not recovered to its pre-event value even after 4 hours following CME onset has elapsed. For EVE \ion{Fe}{14} 211~\AA, the intensity profile shows no dimming feature. This behavior arises because the post-flare loop emissions at high coronal temperatures $\sim$2~MK dominate during the recovery phase \citep{mason14}. 

In the following analysis, we will also find it helpful to screen out the brightening due to post-flare loops in the EVE light curves, and so we make use of a scheme to synthesize the EVE light curves from differential emission measure (DEM) inversion techniques applied to the set of corresponding AIA images. The sparse inversion technique developed by \citet{cheung15} is employed to infer the coronal DEM distribution from 6 AIA channels (94~\AA, 131~\AA, 171~\AA, 193~\AA, 211~\AA, 335~\AA). Synthesized light curves are achieved by convolving the DEM with the contribution function $G(T)$ of EVE lines obtained from CHIANTI database \citep{landi13} to get the synthetic EVE line irradiance. The synthesized curves for the full disk are shown in red in Figure~\ref{eve_obs}, which match the EVE line intensity profiles from the 2011 February 15 event reasonably well. For the DEM analysis, we only use AIA images with the automatic exposure control enabled ({\tt AEC\_TYPE=2}) and the exposure time longer than 0.01 seconds. The AIA instrumental degradation is not considered in this study because this event occurred in early phase of the mission when the degradation was minimal. In Figure~\ref{eve_obs}, the integrated EM in specific temperature bins, as derived from the AIA DEM inversion around the peak emission temperatures of \ion{Fe}{9} 171 \AA~($\log T/K$ = 5.85 to 5.95) and \ion{Fe}{14} 211 \AA~($\log T/K$ = 6.25 to 6.35), are plotted in blue. Both EM evolution profiles show a trend similar to the observed major emission lines in those temperature ranges. After the impulsive phase of the flare, the total plasma content in $\log T/K$ = 5.85 to 5.95 decreases to a lower value than the pre-eruption state, while the plasma content increases in $\log T/K$ = 6.25 to 6.35. We also noticed an increase in EM of $\log T/K$ = 5.85 to 5.95 after reaching the minimum around $\sim$2:30 UT, which could be attributable to plasma cooling shifting emission measure to a lower temperature. This increase is also visible in the EVE \ion{Fe}{9} 171 \AA~irradiance but not as evident as in the EM profile, which could be ascribed to a wider temperature range of plasma that contributes to the \ion{Fe}{9} 171 \AA~irradiance.

\section{Results}
To better understand the coronal dimming evolution, we simulate CME eruptions with different initial flux rope parameters, which lead to a variety of flux rope energies and configurations. This series of cases not only enable us to investigate the coronal dimming characteristics under different CME eruptions but also help to account for the uncertainties of flux rope parameters that are not well constrained from observations. The simulation cases used in this study are summarized in Table~\ref{sim_cases}. Note that 5 out of 7 cases in this study are from our previous study \citep{jin16} and we add two new cases (\texttt{Run 2} and \texttt{Run 4}) in this study. We also calculate the magnetic energy of the inserted flux rope for different cases, which ranges from 4.1$\times$10$^{31}$ erg to 4.2$\times$10$^{32}$ erg. All the simulations are advanced for at least 2 hours when the CMEs are well into the interplanetary space and the global corona has begun to relax. We then synthesize EUV emissions for AIA optically-thin channels and process the synthetic data using the same method as for the observational data (see \S 3). In the following, we summarize the results into three main findings.

\begin{deluxetable}{ccccccc}
\tablecolumns{7}
\tablewidth{0pt}
\tabletypesize{\footnotesize}
\tablecaption{Summary of the Simulation Runs}
\tablehead{
\colhead{}  &  \multicolumn{2}{c}{Flux Rope Parameters} & & \multicolumn{3}{c}{Simulated CME Properties}  \\
\cline{2-3}  \cline{5-7} \\
\colhead{Run Number} & \colhead{$a_1$\tablenotemark{*}} & \colhead{Orientation\tablenotemark{$\dagger$}} & & \colhead{Magnetic Energy [erg]} & \colhead{CME speed [km s$^{-1}$]} & \colhead{CME Mass [g]}} 
\startdata
%1  &  5.0 & 128$^{\circ}$ && 1.1E+31 & 411 & 1.8E+15 \\
1  &  12.5 & 128$^{\circ}$ && 4.1E+31 & 801 & 5.7E+15\\
2 & 12.5 & 90$^{\circ}$ && 4.1E+31 & 921 & 5.7E+15\\
3 & 12.5 & 270$^{\circ}$ && 4.1E+31 & 1216 & 6.5E+15\\
4 & 12.5 & 308$^{\circ}$ && 4.1E+31 & 1277 & 6.4E+15\\
5 & 25.0 & 128$^{\circ}$ && 1.2E+32 & 1598 & 1.5E+16\\
6 & 50.0 & 128$^{\circ}$ && 4.2E+32 & 2607 & 3.7E+16\\
7 & 50.0 & 216$^{\circ}$ && 4.2E+32 & 2668 & 3.9E+16\\
\enddata
\tablenotetext{*}{$a_1$ determines the magnetic strength of the flux rope. The other three parameters of the flux rope are fixed in this study ($a=0.3$, $r_{0}=0.3$, $r_{1}=1.4$), please refer to \citet{jin17b} for the definition of these parameters.}
\tablenotetext{\dagger}{As specified in \citet{jin16}, an orientation angle of 0° means that the foot points of the flux rope are along the east–west direction with the positive polarity at east, while an orientation angle of 180° has the positive polarity to the west. The orientation angle increases in a clockwise fashion.}
\label{sim_cases}
\end{deluxetable}

\subsection{Coronal Dimming Evolution}
\begin{figure}[ht]
\begin{center}$
\begin{array}{c}
\includegraphics[width=1.0\textwidth]{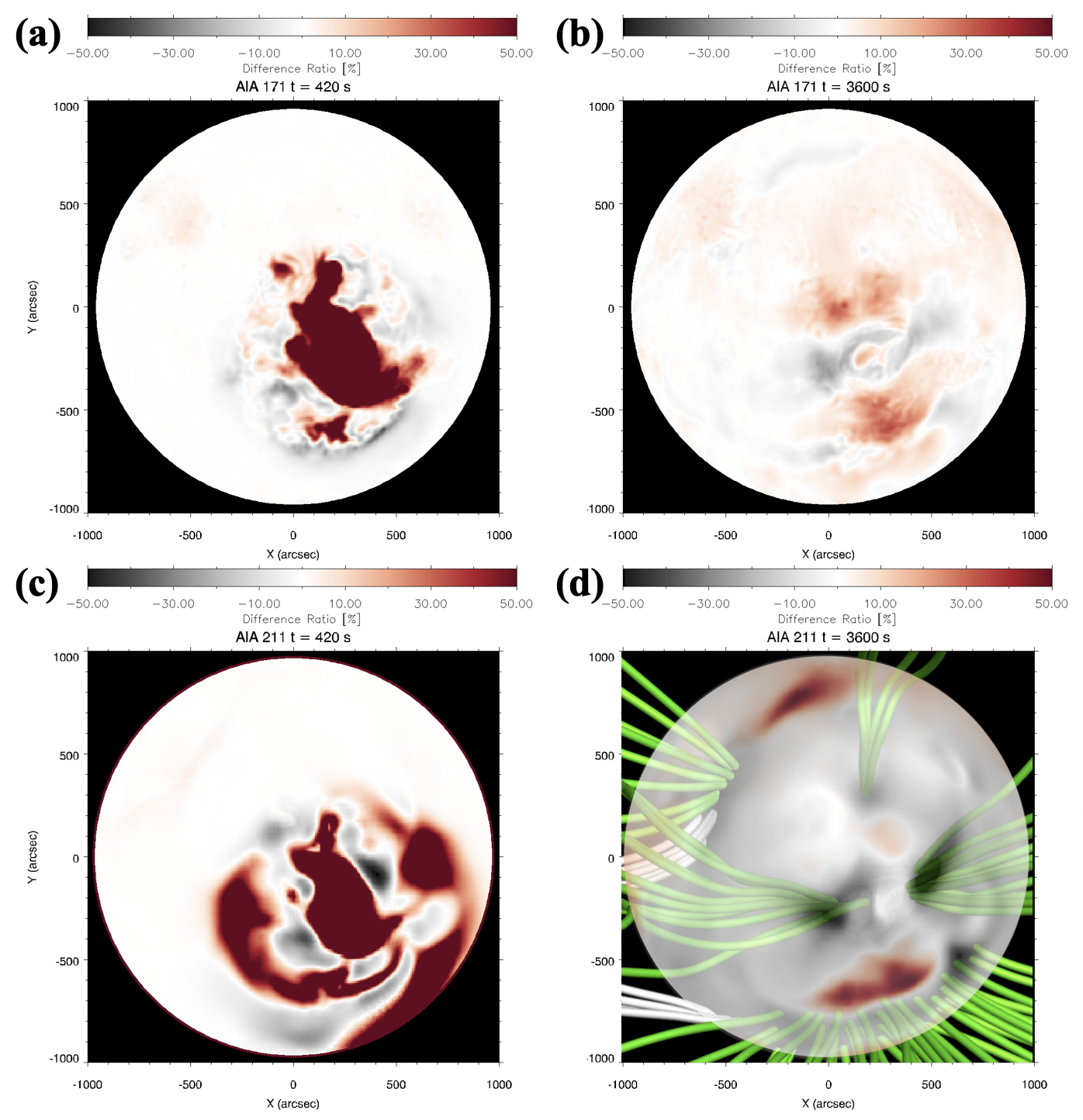}
\end{array}$
\end{center}
\caption{The coronal dimming/brightening simulated by the MHD model (\texttt{Run 1}). Panels (a) and (b) show synthesized AIA 171~\AA~percentage base difference images at $t$ = 7 minutes and 60 minutes. Panels (c) and (d) show synthesized AIA 211~\AA~percentage base difference images at $t$ = 7 minutes and 60 minutes. Selected field lines from dimming regions from \texttt{Run1} are overlaid on panel (d). The green and white field lines represent the open field and large-scale helmet streamer field lines respectively. An animation of this figure is available in the online version of this article, which shows the synthesized AIA percentage base difference movie of 171~\AA~and 211~\AA~in 60 minutes with temporal resolution of 30 seconds.}
\label{aia_sim}
\end{figure}

In Figure~\ref{aia_sim}, the simulated coronal dimming/brightening associated with the CME eruption is shown for \texttt{Run 1} at two different times $t$ = 7 minutes and $t$ = 60 minutes. Synthesized percentage base-difference images for the same AIA channels (171~\AA~and 211~\AA) used for Figure~\ref{aia_obs} are shown in Figure~\ref{aia_sim}. The simulation reproduced several key features of the observation: (1) the darkening/brightening waves in the low/high temperature AIA channels, (2) the core dimmings from the source active region in both channels, and (3) remote dimmings away from source region (e.g., area around [X, Y] = [-500\arcsec, 300\arcsec]) occurring during the recovery phase. 

\begin{figure}[ht]
\begin{center}$
\begin{array}{c}
\includegraphics[width=1.0\textwidth]{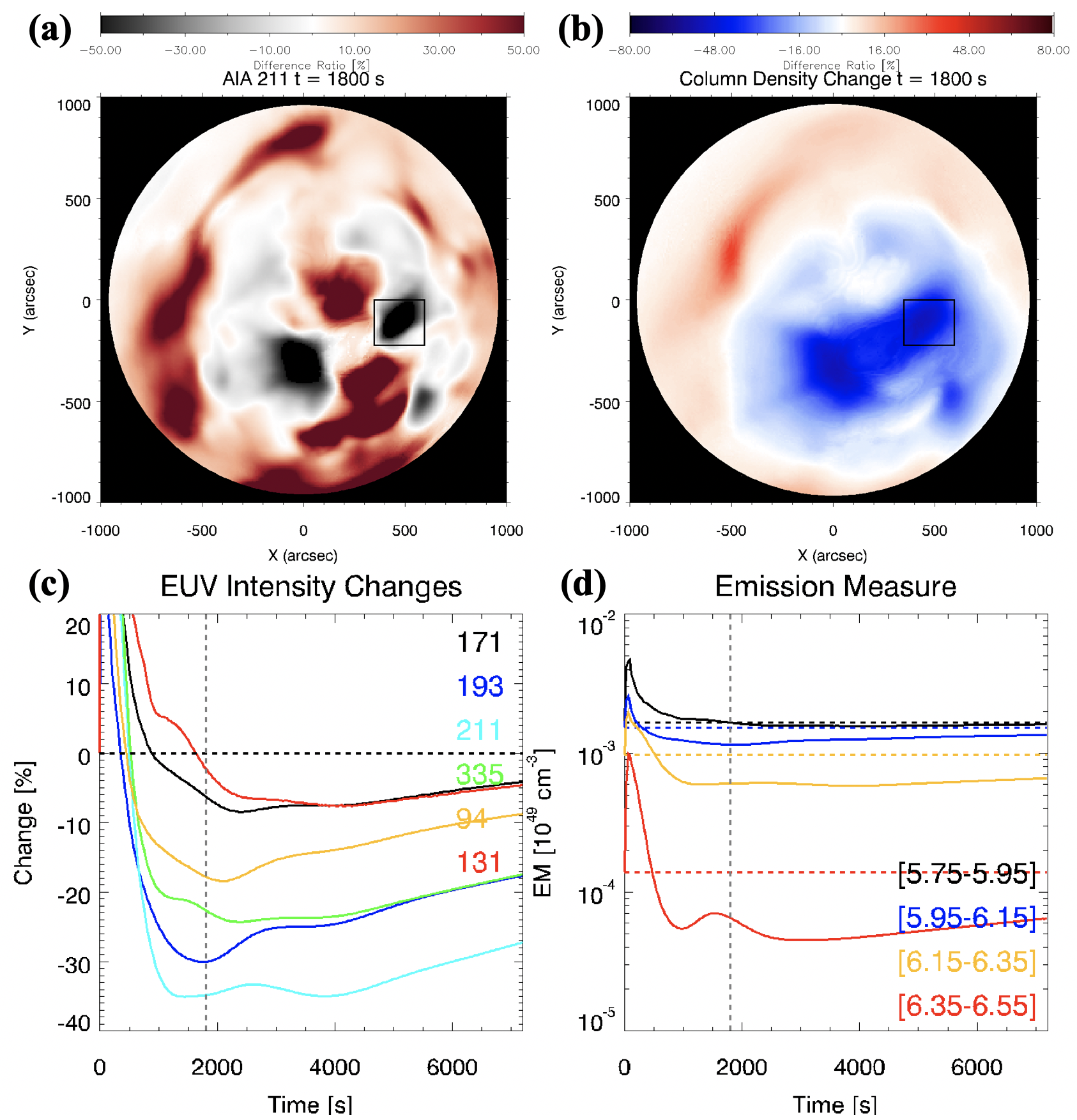}
\end{array}$
\end{center}
\caption{Dimming region evolution in the simulation (\texttt{Run 1}). (a) Synthesized AIA 211~\AA~percentage base difference image at $t$ = 30 minutes. The black box shows the sub-region where the EUV intensity and Emission Measure (EM) are derived. (b) LOS integrated column density change at $t$ = 30 minutes. (c) The EUV intensity changes in six synthesized AIA channels. (d) EM evolution in four temperature bins from $\log T/K$ = 5.75 to 6.55 in the simulation. The horizontal dashed lines show the pre-event intensity and EM values. The vertical dashed lines show the timing of (a) and (b).}
\label{core_dimming}
\end{figure}

To understand the plasma evolution associated with the coronal dimming, we extract the EUV intensity evolution in 6 AIA channels for selected sub-regions. In addition, we calculate the evolution of the emission measure (EM) in four different temperature ranges from $\log T/K$ = 5.75 to 6.55, which covers the temperature range in which plasma contributes to the coronal dimming regions in this study. In Figure~\ref{core_dimming}, the core dimming region evolution is shown for \texttt{Run 1}. The black box in the percentage base-difference image in panel~(a) shows the sub-region where the EUV intensity (Figure~\ref{core_dimming}(c)) and EM profiles (Figure~\ref{core_dimming}(d)) are derived. We can see that after initial brightening due to the flare, all synthetic AIA line intensities decrease below pre-flare values. The AIA 211~\AA~ intensity decreases the most (by $\sim$35\%) and the 131~\AA~decreases the least (by only $\sim$5\%). The EM profiles also suggest plasma depletes in all temperature bins from $\log T/K$ = 5.75 to 6.55. Furthermore, to confirm the core dimming is due to the actual mass loss instead of other physical processes (e.g., temperature changes of stratified solar atmosphere), we also calculate integrated line-of-sight (LOS) column density change, which is shown in Figure~\ref{core_dimming}(b). The column density is defined as $N=\int n_{e} dl$, where $n_{e}$ is the electron density, and $dl$ is a path length along the LOS. The result demonstrates that the core dimming is caused by CME-induced plasma depletion. Note that the intensity changes in the recovery phase are quasi-linear for all AIA channels. Based on linearly extrapolating the intensity curves, we estimate that the dimming will be recovered in $\sim$9-16 hours after the CME onset, which is consistent with the EVE \ion{Fe}{9} 171 \AA~observation (the irradiance recovers to the pre-event value $\sim$10:00 UT) as well as the statistical coronal dimming studies \citep{reinard08}.

\begin{figure}[ht]
\begin{center}$
\begin{array}{c}
\includegraphics[width=1.0\textwidth]{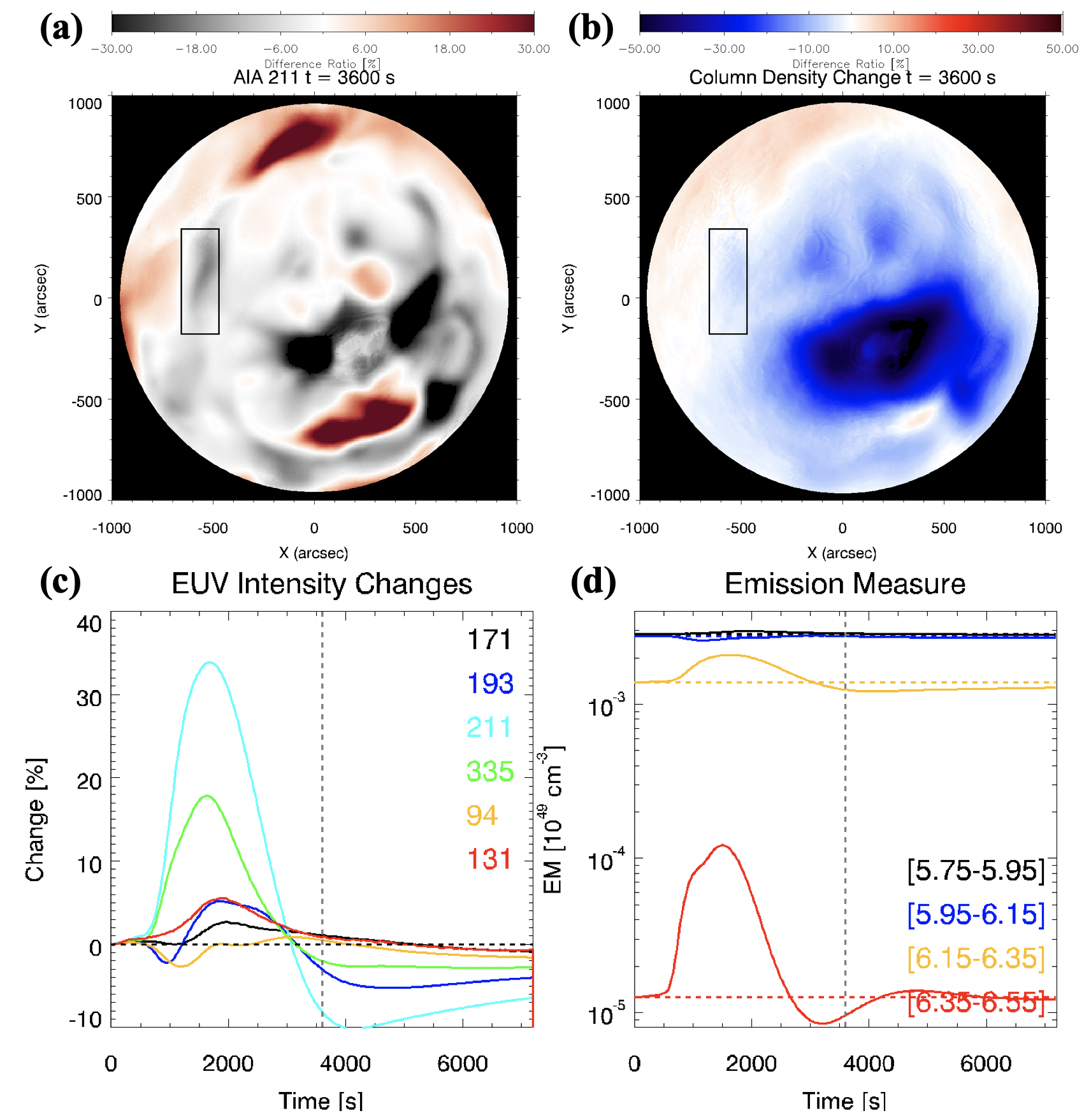}
\end{array}$
\end{center}
\caption{The same as Figure~\ref{core_dimming} but for a different sub-region and at a later time.}
\label{remote_dimming}
\end{figure}

\begin{figure}[ht]
\begin{center}$
\begin{array}{c}
\includegraphics[width=1.0\textwidth]{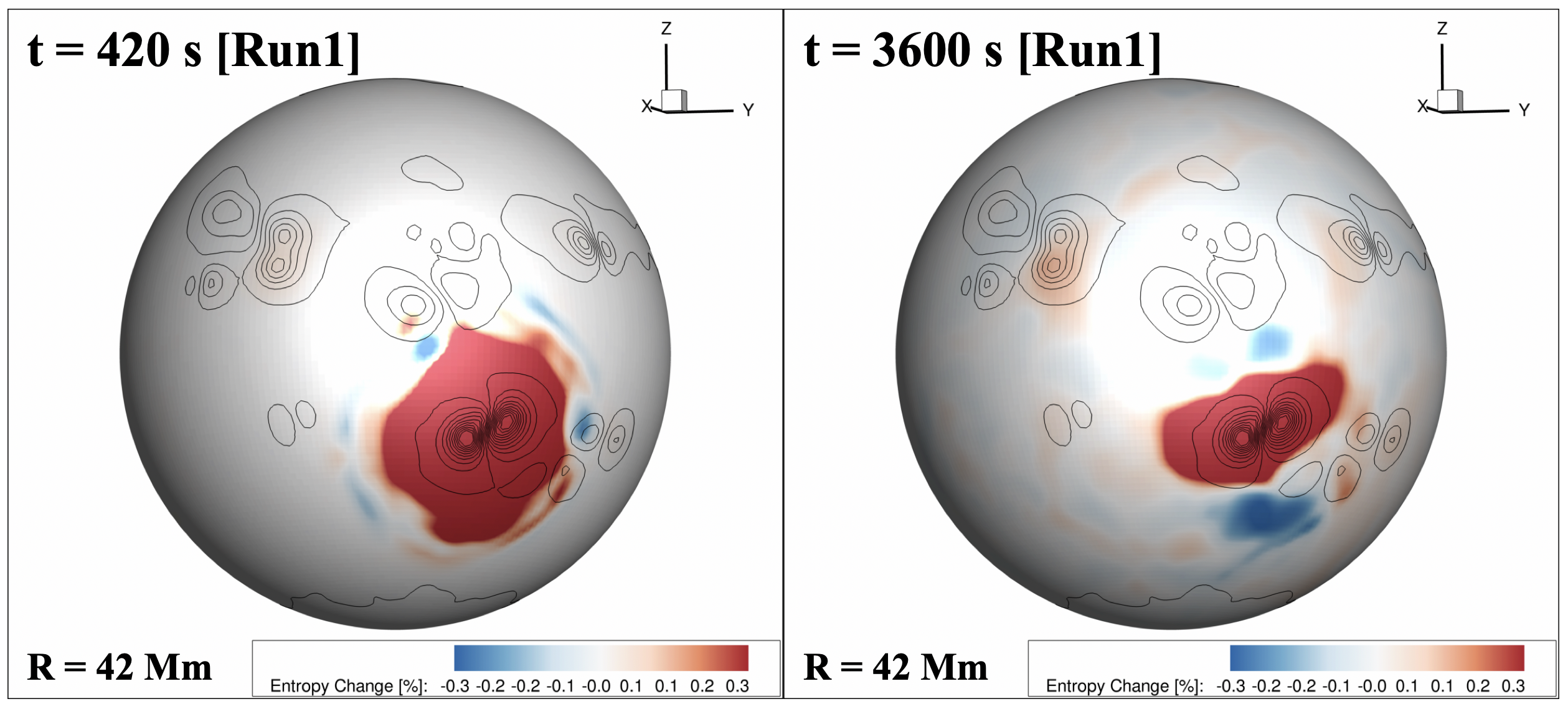}
\end{array}$
\end{center}
\caption{Entropy change after 7 minutes (left panel) and 60 minutes (right panel) in the simulation over an isosurface at a height of 42 Mm above the model photosphere. The simulation data is from \texttt{Run 1}. The black contours are of the radial magnetic field and indicate different active regions.}
\label{entropy}
\end{figure}

A remote dimming area near the east limb developed later in the evolution, which is evident both in the observation (see Figure~\ref{aia_obs}) and simulation (see Figure~\ref{aia_sim}). In Figure~\ref{remote_dimming}, the EUV intensity and EM profiles are shown for that region. For this region of interest near the east limb, the synthetic AIA intensity curves evolve in a different fashion than for the core. Several lines (e.g., 211, 335~\AA) show increased intensity at the early-stage while the other lines (e.g., 193, 94~\AA) show decreased intensity at the same time. But the physical process is better demonstrated with the EM profiles, in which we can see the plasma density decreases in lower temperature bins and increases in higher temperature bins. This evolution is consistent with the plasma being heated to higher temperatures due to the adiabatic compression during the CME eruption. Figure~\ref{entropy} illustrates the specific entropy change, in which we calculated the entropy change at the height of 42 Mm with $t$ = 7 minutes and $t$ = 60 minutes in the simulation case \texttt{Run 1}. The specific entropy is defined as $s=\ln(T/\rho^{\gamma-1})$, where $T$ is the plasma temperature and $\gamma$ is the polytropic index {($\gamma$=5/3)}. The specific entropy is invariant for plasma undergoing adiabatic compression/expansion, but can change due to nonadiabatic processes such as thermal heat conduction, and Ohmic and viscous dissipation \citep{cheung07}. The black lines in the figure correspond to contours of the radial magnetic field, and indicate locations of the on-disk active regions. By comparing the EUV percentage base-difference images in Figure~\ref{aia_sim}a and \ref{aia_sim}c with the entropy change shown in the left panel of Figure~\ref{entropy} at the same time ($t$ = 7 minutes), it is evident that the entropy mainly changes within an area near the source region where magnetic reconnection occurs, i.e., between the erupting flux rope and the surrounding coronal field. However, the EUV wave traverses locations where there is small or no entropy change, which is more indicative of adiabatic temperature changes. At $t$ = 60 minutes, the region where entropy has changed is now smaller than that at $t$ = 7 minutes, as the flux rope has lifted off into the heliosphere and magnetic reconnection in the corona tapers off. The enhanced entropy is mainly due to the mass loss as shown in Figure~\ref{remote_dimming}(b). There is also an area with decreased entropy to the south of the source active region, which is likely related to the density increase seen in the column density plot in Figure~\ref{remote_dimming}(b). In addition, based on the column density change, this remote dimming (which developed later in the evolution) is also caused by the mass loss, although the magnitude is much less than the core dimming area.

\subsection{Dimming under Different Flux Rope Orientations}
\begin{figure}[ht]
\begin{center}$
\begin{array}{c}
\includegraphics[width=1.0\textwidth]{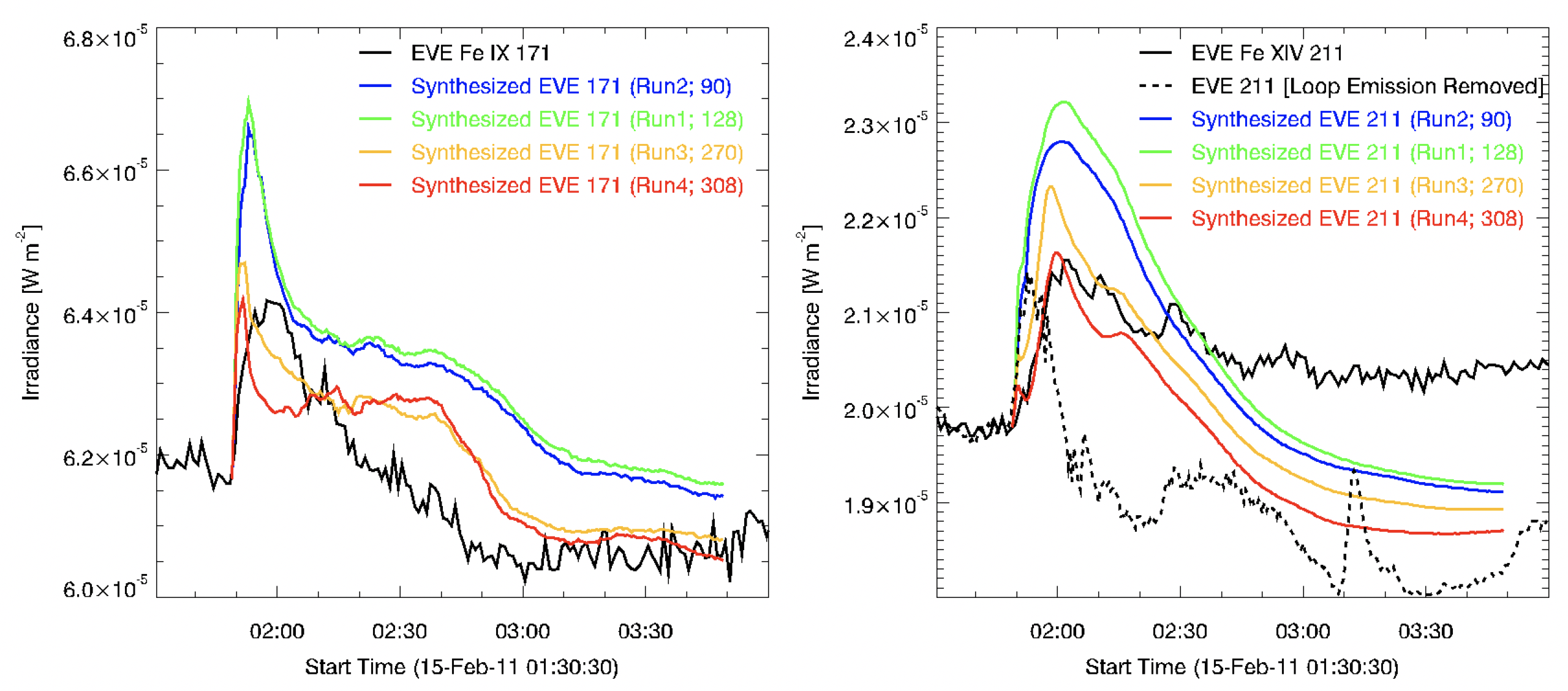}
\end{array}$
\end{center}
\caption{Comparison between EVE observation and synthesized EVE line intensity in the simulations \texttt{Run1} -- \texttt{Run4}. The numbers in the brackets indicate the flux rope orientation angles as shown in Table~\ref{sim_cases}. Note that the pre-event value of the synthetic line intensities are shifted to match the EVE intensities. The dashed line in the EVE 211 figure shows the intensity evolution in the selected area marked in Figure~\ref{aia_obs} with post-flare emission removed.}
\label{syn_eve}
\end{figure}

As discussed in \citet{jin16}, the orientation of the flux rope in the simulation has a significant influence on the interaction between the flux rope and the surrounding corona. Here, we investigate how the coronal dimming properties vary when the initial flux rope orientation is varied. Simulation cases \texttt{Run 1} -- \texttt{Run 4} listed in Table 1 form a set of simulations in which the flux ropes in these simulations possess the same initial magnetic energy but have different orientations. We run each of the four simulation cases for two hours and synthesized two EVE light curves (\ion{Fe}{9} 171 \AA~and \ion{Fe}{14} 211 \AA) for comparison. The plots in Figure~\ref{syn_eve} show the four simulation cases overlaid with the EVE observation. In the figure, the steady-state synthetic line irradiances are scaled (by 4.4 in 171~\AA~and 16.5 in 211~\AA) to match the pre-event EVE observations. There are multiple reasons for this discrepancy: first, the magnetic map used in the model is fixed at the inner boundary for getting the steady-state solution. Therefore, it may not capture the emissions caused by the continuously evolving solar atmosphere (e.g., flux emergence and dynamics associated with it, coronal jets, etc.). Second, being a global simulation, the spatial resolution is not high enough to resolve fine active region structures. We find that the simulations do reproduce the global coronal state reasonable well (see a quantitative comparison case in \citealt{jin17a}), and because in this work we use relative differences to investigate the dimming properties, the pre-event model discrepancy is not expected to significantly alter the main conclusions.

In Figure~\ref{syn_eve}, the synthetic line intensities show different evolution profiles in the four cases, with \texttt{Run 1 \& 2} having a higher intensity peak but weaker dimming feature while \texttt{Run 3 \& 4} having a lower intensity peak but stronger dimming feature. Considering that all four cases are started with the same initial flux rope energy, a higher peak intensity is indicative of more magnetic reconnection between the flux rope and surrounding coronal field, which leaves less free energy available to accelerate the CME. As a result, the coronal dimming appears weaker.

For this event, the EVE \ion{Fe}{14} 211 \AA\ line irradiance does not show a dimming signal at all, and instead brightens after the eruption. However, in the spatially resolved AIA 211~\AA~ observations, some dimming is evident (see Figure~ \ref{aia_obs}). We interpret the EVE \ion{Fe}{14} 211 \AA\ brightening as primarily due to post-flare loop emissions from the flare source region, which is caused by chromospheric evaporation occurring after the impulsive phase of the flare \citep{woods11}. During this process, the chromospheric plasma is heated by energetic particles and flows up into the active region loops that leads to enhanced post-flare loop emissions. However, this physical process is not captured in our simulation as the model has a fixed density at the inner boundary. Therefore, the synthetic EVE \ion{Fe}{14} 211~\AA~intensity shown in Figure~\ref{syn_eve} is dominated by the dimming signal. In order to evaluate post-flare loop emissions, we select an area (marked as black box in Figure~\ref{aia_obs}) covering the source active region but remove a sub-region with significant post-flare emissions (marked as a dashed box in Figure~\ref{aia_obs}). We then extract the EVE \ion{Fe}{14} 211~\AA~irradiance synthesized from AIA DEM inversion for this selected area. The temporal evolution profile (dashed line) is overlaid in the right panel of Figure~\ref{syn_eve} (scaled to EVE pre-event value). After removing the post-flare emission in 211~\AA~intensity, the dimming signal associated with the CME eruption is revealed to have a similar magnitude as shown in the simulation. Note that the small spike $\sim$03:10 in the dashed line is due to a coronal jet activity from the east edge of the active region, which is overwhelmed by the intense emission from the post-flare loop in the full-disk integrated EVE light curve but stands out after the post-flare loop region emission is removed.

\subsection{Relationship between Dimming Slope/Depth and CME Speed/Mass}
\begin{figure}[ht]
\begin{center}$
\begin{array}{c}
\includegraphics[width=1.0\textwidth]{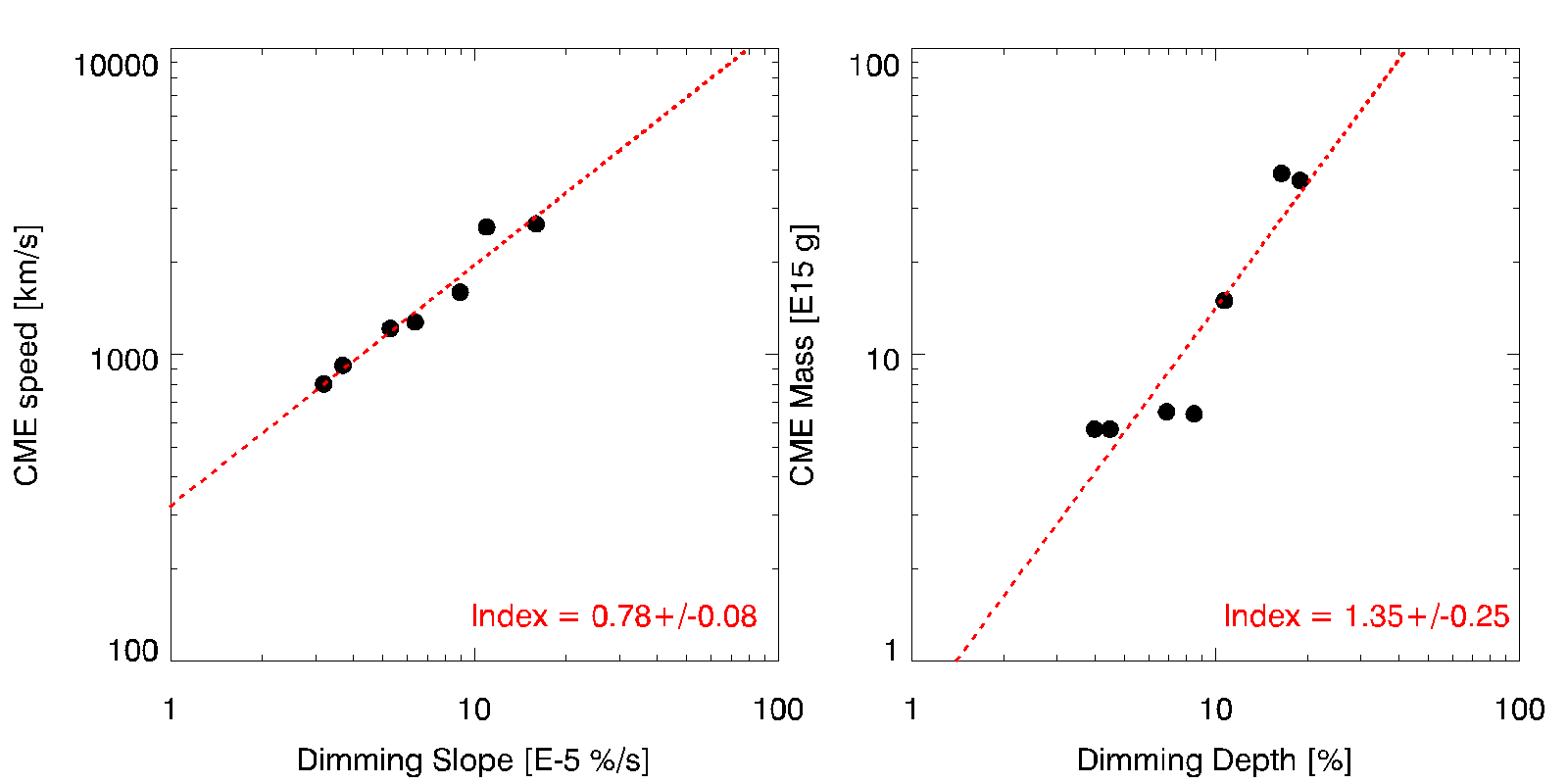}
\end{array}$
\end{center}
\caption{Left panel: relationship between the CME speed and the dimming slope for all 7 simulation cases. Right panel: relationship between the CME mass and the dimming depth for all 7 simulation cases.}
\label{cme-dimming}
\end{figure}

In this section, we use all 7 cases from Table 1 to investigate the relationship between coronal dimming and CME characteristics. To compare with the observational result by \citet{mason16}, we choose the same set of parameters to characterize CME and coronal dimming. For CME, two critical parameters (CME mass and CME speed) are derived for all 7 simulation cases. The CME speed is derived as the average speeds of the CME at the outermost front between 20 and 30 minutes, which is the same determination used in \citet{jin17b}. The outermost front of the CME is determined by finding the CME propagation plane and extracting the line profiles along the CME propagation path (see \citealt{jin13} for an example). The CME mass is calculated as the integrated mass difference in the simulation domain larger than 2 R$_{\odot}$ between the pre-event time $t_{0}$ and at $t_{1}$ = 30 minutes: $m_{CME}= \int_{r=2 R_{\odot}}^{r=24 R_{\odot}} \rho(t_{1})dV - \int_{r=2 R_{\odot}}^{r=24 R_{\odot}} \rho(t_{0})dV$, where 24 R$_{\odot}$ is the outer boundary of the simulation domain.  For coronal dimming, the dimming slope and dimming depth are derived from the simulations using the same definition as \citet{mason16}. In addition, we use synthesized the EVE \ion{Fe}{14} 211 \AA~ line irradiancce when calculating the dimming parameters because for some simulation cases, EVE \ion{Fe}{9} 171~\AA~ dimming is not very evident. The results are shown in Figure~\ref{cme-dimming}. The figure shows that a similar relationship between the CME speed/mass and dimming slope/depth is found for the simulations presented here as in the EVE observations \citep{mason16}. The tight relationship between CME and dimming parameters for the set of simulation cases is likely related to the fact that the modeled CME erupts from the same active region into the same background solar wind configuration, whereas the analogous relationship determined from observations involves an ensemble of CME events from multiple active regions. We also found that although this relationship seems to hold well for the CMEs with speeds larger than $\sim$800 km s$^{-1}$ and up to $\sim$2700 km s$^{-1}$, it could break for slow CMEs. For example, we analyze the \texttt{Run 11} case in \citet{jin16}, in which the initial flux rope has an energy of 1.1$\times$10$^{31}$ erg that leads to a CME speed of 411 km s$^{-1}$. Although the dimming is still evident in the spatially resolved synthetic AIA images, it is not shown in the synthetic EVE light curves. Nevertheless, the simulation reproduces the general trend between the CME and dimming parameters discovered in the observations that stronger CMEs result in greater dimming signals.

\section{Discussion \& Conclusions}
\begin{figure}[ht]
\begin{center}$
\begin{array}{c}
\includegraphics[width=1.0\textwidth]{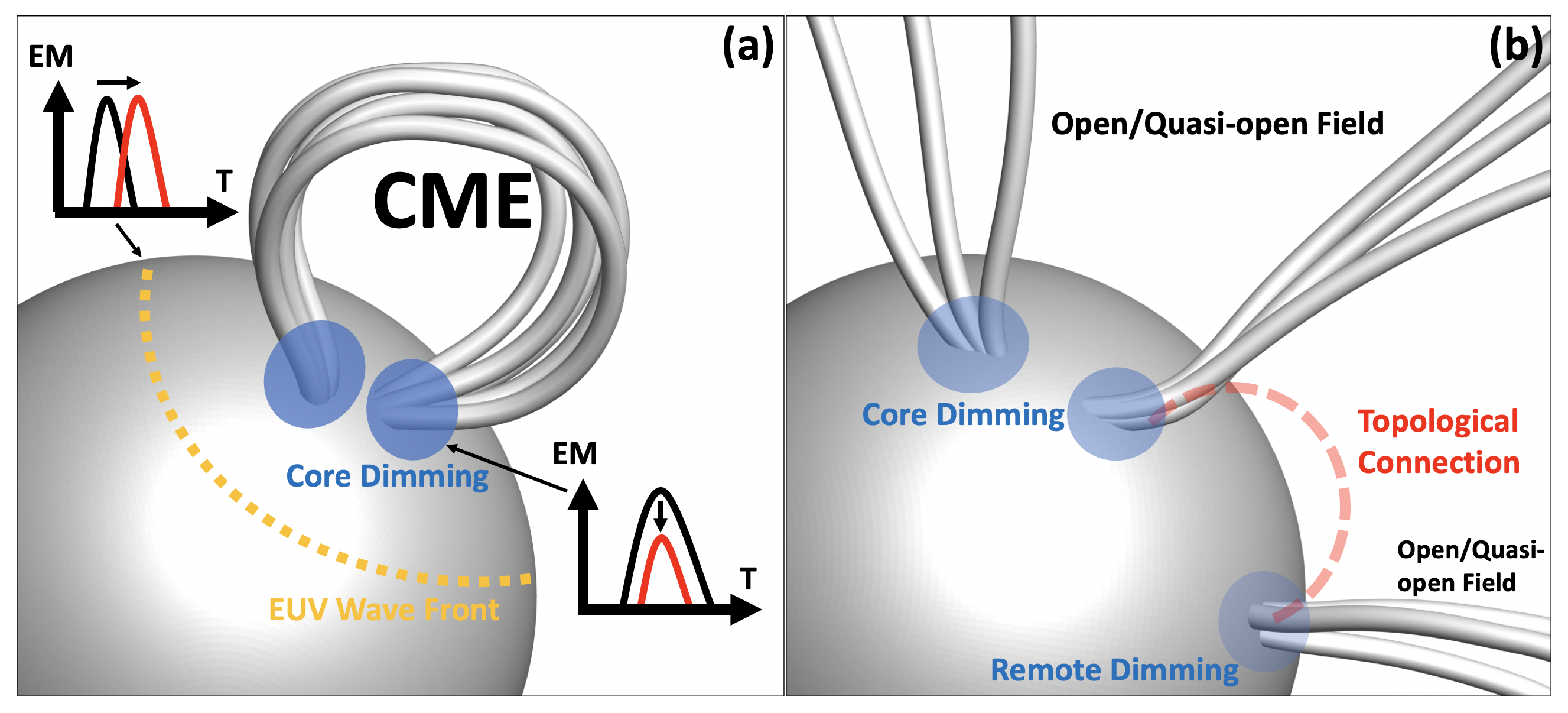}
\end{array}$
\end{center}
\caption{A cartoon figure showing the relationship between the CME, coronal dimmings, and EUV waves (a) at the early-stage and (b) at a later phase. The black and red curves in the inset plots of EM vs. T represent the initial and disturbed plasma states, respectively.}
\label{cartoon}
\end{figure}

In this study, by combining SDO observations and advanced MHD modeling, we conducted a comparative study of the coronal dimming induced by the CME on 2011 February 15. The results show that our model could reproduce many observed features of the coronal dimming after the CME eruption. By varying the flux rope properties, the ensemble of models possess a close relationship between the CME and dimming characteristics, similar to the trends found from observations of coronal dimming events \citep{mason16, dissauer18, dissauer19}. Our result suggests that coronal dimmings encode important information about CME energetics, CME mass, and the magnetic configuration of erupting flux ropes. Moreover, with the help of DEM and column density analysis of the simulation data, we found that the transient dimming / brightening patterns are related to plasma heating processes (either by adiabatic compression or reconnection), while the long-lasting ``core" and ``remote" dimmings are caused by mass loss process induced by the CME eruption. We illustrate the relationship between the CME, dimmings, and EUV waves in Figure~\ref{cartoon}.

We found that when relating the long-lasting dimming patterns across the solar disk with the global coronal field configuration (as shown in Figure~\ref{aia_sim}d), the dimming areas are always associated with open or quasi-open field lines. We found that most of these open/quasi-open field lines are pre-existing, i.e., they are present in the steady-state solution before the CME eruption (see Figure 5 in \citealt{jin16}). But some of the open field in the CME source region is formed after the eruption due to the magnetic reconnection between the flux rope and surrounding coronal field as it propagates out into the heliosphere. This is consistent with the view that the core dimmings correspond to the footpoints of the erupting flux rope system but how the remote area far away from the CME source region loses mass that eventually leads to long-last dimmings? By further comparing the remote dimming areas with the global topological structures of the corona (see Figure 13 in \citealt{jin16}), we found that all the remote dimming areas are connected to the source region by some topological structures (e.g., separatrix surfaces) so that the plasma evacuation in the source region could directly influence these remote areas. On the other hand, it is the open/quasi-open field in these regions that provides a pathway for the plasma to leave the Sun, and as a result these remote dimming patterns can exist for hours after the CME eruption. In sum, both connection to the source region and the open/quasi-open field are required for getting the remote dimmings. 

Furthermore, the CME-dimming relationship could also be important for space weather forecasting purposes because the CME-induced coronal dimming (especially the core dimming) occurs much earlier than other observable CME disturbances. Currently, we are able to estimate the CME speed only when it is observed in the white light coronagraph at least 20-30 minutes after the CME onset. With a reliable CME-dimming relationship, it is possible to estimate the CME speed at an earlier phase of the eruption. This information is important for most space weather forecasting models, especially those forecasting the propagation solar energetic particle models as the SEP can arrive at Earth in minutes to hours after CME onset. Additionally, for the type of modeling presented here, the EEGGL model \citep{jin17b} requires an estimate of the CME speed in setting the flux rope parameters, and consequently early estimates of the CME speed will better constrain EEGGL and improve the model performance.

As mentioned \S 1, \citet{schrijver11} studied the same 2011 February 15 event with detailed observation and MHD modeling efforts. Therefore, it would be interesting to compare the results from the two individual studies about the same event. There are many aspects that the two studies agree on for the event. For example, the adiabatic nature of the expansion front as discussed in \S 4.1, as well as the non-adiabatic process that is needed to explain the northward-propagating segment of the front. In \citet{jin16}, we found that this non-adiabatic process is caused by the magnetic reconnection between the expanding flux rope and the active region field to the north of the CME source region. Also, in both MHD models presented in the studies, the wave-like warming compression front moves significantly ahead of the erupting flux rope, which is also shown in other recent MHD simulations of EUV waves (e.g., \citealt{mei20, downs21}). However, there are also differences found in the two models. In this study, we found two distinct wave-like structures (as shown in Figure 10 of \citealt{jin16}): first, the outmost one (i.e., warming compression front) demonstrates more features like a fast-mode wave as it does not stop at the topology boundaries (e.g., the helmet structures mentioned in \citealt{schrijver11}). However, it is evidently weakened passing these boundaries likely due to the wave reflection. The second structure is related to the expanding flux rope volume. As discussed in \citet{jin16}, the erupting flux rope does not easily break out from the topological structures therefore stopped by the streamer structures. Note that the flux rope and background coronal field settings are quite different in these two simulations, which could largely contribute to the differences mentioned above. We will leave the more detailed comparison for future study as we concentrate more on the coronal dimmings induced by the CME in this study. 

Finally, we discuss the possibility of using coronal dimming to detect stellar CMEs. Even though the stellar flares are frequently observed, it remains challenging to associate stellar CMEs with any of these flares. The ability to distinguish between stellar flares and CME events is important since the two phenomena affect exoplanet habitability through very different physical processes (as is true in our own solar system). CMEs have a much larger influence than flares on stellar evolution through mass loss and angular-momentum loss (e.g., \citealt{benz10}), and can significantly impact the habitability of exoplanets due to their ability to erode their atmospheres (e.g., \citealt{lammer07, airapetian20}). \citet{argiroffi19} detected a stellar flare with blue-shifted \ion{O}{8} line with 90$\pm$30 km s$^{-1}$ on the active star HR 9024 using data from Chandra X-ray Observatory space telescope, which might indicate a CME eruption in place. However, this kind of observation is still very close to the star surface and therefore reflects the very early-stage of the eruption. It is hard to obtain signals at greater heights, and as a result we do not know whether the eruption propagated into the astrosphere, and, if so, what its speed and mass are. Indeed, MHD modeling  suggests that confined eruptions could be common for stellar case due to the stronger magnetic confinement \citep{alvarado18}. Additionally, Type II radio bursts caused by accelerated electrons associated with CME-driven shocks seem to be conspicuously absent \citep{crosley18a, crosley18b}. More recently, using H$_{\alpha}$ spectroscopic observations,   \citet{namekata21} detected a strong blueshifted absorption component with a velocity of $\sim$510 km s$^{-1}$ associated with a TESS white-light flare on the young solar-type star EK Draconis, which presents a probable detection of an eruptive filament. A comparative study with the solar events further suggests that this filament likely leads to a CME into the interplanetary space.

The ability to use coronal dimming signatures as a proxy for detecting stellar CMEs seems promising, and has been recently explored in more detail by \citet{veronig21} using historical data from XMM, HST, and EUVE, from which they identified 21 stellar dimming events (i.e., CME candidates) on 13 different stars. The statistics of these dimming events suggest stronger dimming compared with the solar case, with the strongest stellar dimming event corresponding to a depletion of half of the stellar corona. In another study, \citet{jin20} modeled the stellar dimming with enhanced magnetic flux density for M dwarf and young Sun-like stars and found that, due to the stronger magnetic field and coronal heating, the spectral lines showing stellar dimmings are shifted to higher temperature range. Although it remains possible that stellar dimming patterns may differ in significant ways from the solar case, coronal dimming phenomena are likely ubiquitous phenomenon for all G-M type stars. And for the stars with a similar mass, temperature, and magnetic field strength as the Sun, solar coronal dimming events, such as the one shown in this article, serve as good references for comparative studies of stellar coronal dimming. In addition, by applying the instrument performance estimates from the Extreme-ultraviolet Stellar Characterization for Atmospheric Physics and Evolution (ESCAPE; \citealt{france22}), which will provide extreme- and far-ultraviolet spectroscopy (70 - 1800~\AA), the study demonstrates that, with instrumentation optimized for these measurements, we would be able to detect the stellar coronal dimming in UV/EUV range. Meanwhile, current and further solar observations (e.g., MUSE: Multi-slit Solar Explorer; \citealt{bart20, bart21, cheung21}) will keep facilitating our understanding about solar coronal dimming events, their relationship to CME, and their applications to solar-stellar connection studies.

\begin{acknowledgements}
MJ, MCMC, MLD, NVN are supported by NASA's SDO/AIA contract (NNG04EA00C) to LMSAL. MJ thank the International Space Science Institute (Bern, Switzerland; https://www.issibern.ch) for supporting the Working Team 516 on ``Coronal Dimmings and Their Relevance to the Physics of Solar and Stellar Coronal Mass Ejections" led by Astrid Veronig \& Karin Dissauer. MJ also thank Cooper Downs at Predictive Science Inc. and James Mason at JHU/APL for helpful comments and discussions. We thank the NASA’s Living With a Star Program, which SDO is the first flagship mission of, with AIA, HMI, and EVE instruments on-board. CHIANTI is a collaborative project involving George Mason University, the University of Michigan (USA), University of Cambridge (UK) and NASA Goddard Space Flight Center (USA). The simulation results were obtained using the Space Weather Modeling Framework (SWMF), developed at the Center for Space Environment Modeling (CSEM), University of Michigan (\url{https://github.com/MSTEM-QUDA/SWMF}). We are thankful for the use of the NASA Supercomputer Pleiades at Ames and its helpful staff for making it possible to perform the simulations presented in this paper. 
\end{acknowledgements}

\newpage
\bibliographystyle{aasjournal}
%\bibliography{ref}

\end{document}